\documentstyle[newarcrc,fleqn]{article}
\input{epsf.sty}

\def\plottwo#1#2{\centering \leavevmode
\epsfxsize=.45\columnwidth \epsfbox{#1} \hfil
\epsfxsize=.45\columnwidth \epsfbox{#2}}

\newcommand{\msun}{M_\odot}
\newcommand{\dmc}{\dot M_{\rm cr}}
\newcommand{\dmej}{\dot M_{\rm ej}}
\newcommand{\myr}{M_\odot {\rm yr}^{-1}}
\newcommand{\be}{\begin{equation}}
\newcommand{\ee}{\end{equation}}
\def\ueber#1#2{{\setbox0=\hbox{$#1$}%
  \setbox1=\hbox to\wd0{\hss$ #2$\hss}%
  \offinterlineskip
  \vbox{\box1\box0}}{}}
\newcommand{\la}{\; \lower 1mm \hbox{\ueber{\sim}{<}} \;}
\newcommand{\ga} {\; \lower 1mm \hbox{\ueber{\sim}{>}} \;}

\title{Binary evolution with relativistic jets}

\author{Ulrich Kolb\address{Astronomy Group, University of Leicester, 
        Leicester LE1 7RH, U.K.}
        \thanks{also Max--Planck--Institut f\"ur Astrophysik,
        85740 Garching, Germany}}

\begin{document}
% typeset front matter
\maketitle

\begin{abstract}
Relativistic jets can extract mass--energy from a black hole. In
semi--detached black hole binaries the jet ejection process
constitutes a `consequential angular momentum  loss' (CAML)
process. The effect of this jet--induced CAML is to lower  
the transfer rate below the value set by systemic driving and 
to stabilize otherwise unstable systems.
Implications of jet--induced CAML for GRO~J1655-40 are 
discussed.   
\end{abstract}

\section{Introduction}

Soft X--ray transient outbursts are widely believed to represent
the bright states of an accretion disc limit--cycle, caused by the
presence of partially ionised hydrogen in the disc. For this
instability to operate the mass supply rate into the disc must be less
than a critical value $\dmc$ which would keep the disc just hot enough
for  
hydrogen being always fully ionised. The dominant role of disc
irradiation in systems with a neutron star or black hole primary 
makes $\dmc$ much smaller than in systems with a white dwarf accretor
(van~Paradijs 1996, King et al.\ 1996). 

The black hole X--ray transient source GRO~J1655--40 represents a challenge
for this picture as the donor's observed location in the HR diagram ---
in the middle of the Hertzsprung gap (Orosz \& Bailyn 1997, van der
Hooft et al.\ 1998) --- implies a transfer rate $\simeq 5 \times 10^{-7}
\myr$, which is about a factor of 10 larger than  
$\dmc$ (Kolb et al.\ 1997, Kolb 1998).  
To resolve this discrepancy it has been suggested that 
the donor is either in a short--lived phase where the expansion through
the Hertzsprung gap slows (Kolb et al.\ 1997), or still on the
main sequence, which could be widened by convective overshooting in the star  
(Reg\H os et al.\ 1998).  

Here we investigate a third possibility, stimulated by the observation
that GRO~J1655--40 is a relativistic jet source which may harbour a 
black hole close to maximum spin (Zhang et al.\ 1997, Cui et
al.\ 1998). We show that the continuous ejection of a 
highly relativistic jet, powered by the black hole rotational energy,
decreases the evolutionary mean mass transfer rate.

\section{Mass transfer stability}

The mass transfer rate in a semi--detached binary with accretor mass
$M_1$ and donor mass $M_2$ is determined by the difference between the
donor's stellar radius $R$ and its Roche lobe 
radius $R_L$. Stationary mass transfer implies that star and lobe move
in step, hence $\dot R/R = \dot R_L/R_L$. The radius change can be
written as  
\be
  \frac{\dot R}{R}  = \zeta \frac{\dot M_2}{M_2} + K 
\label{r}
\ee
(e.g.\ Ritter 1996). 
Both the stellar (local) mass--radius index $\zeta$ and the
logarithmic radius change $K$ in the absence of mass transfer (e.g.\ 
nuclear expansion) are determined by the donor's structure alone. 
In principle they are known from stellar structure
calculations. Typically $\zeta$ varies from $-1/3$ to $1$.
The Roche lobe radius changes as 
\be
  \frac{\dot R_L}{R_L}  = \zeta_L \frac{\dot M_2}{M_2} + 2 \frac{\dot
       J_{\rm sys}}{J} ,
\label{rl}
\ee
where $\zeta_L$ is the Roche--lobe index and $J$ 
the orbital angular momentum. 
The systemic term $\dot J_{\rm sys}$ refers to angular momentum
losses with no (or negligible) mass loss from either 
component, e.g.\ gravitational wave emission. The Roche--lobe index
can be obtained from the time derivative of 
$J=M_1 M_2 (Ga/M)^{1/2}$ ($a$ is the orbital separation),  
by noting that $R_L=f_2 a$, with a mass ratio--dependent geometry
factor $f_2$. 
In the case of conservative mass transfer ($J$ and total binary mass $M$
constant) $\zeta_L = \zeta_{L0} = 2M_2/M_1 - 5/3$ if $M_2 \la M_1$. 

Equating (\ref{r}) and (\ref{rl}) and solving for $\dot M_2$ gives the
familiar expression
\be
     \frac{\dot M_2}{M_2} = \frac{2\dot J_{\rm sys}/J - K}
          {\zeta-\zeta_L}
\label{mdot}
\ee
for the stationary mass transfer rate, consisting of the systemic driving 
(numerator) and the stability term (denominator). A simple stability
analysis (e.g.\ King \& Kolb 1995) gives $\zeta-\zeta_L > 0$ as a
necessary condition for stable mass transfer. In most cases $\zeta_L$
increases with mass ratio $q=M_2/M_1$, hence the stability
limit places an upper limit on $q$.
The transfer rate in unstable systems 
would grow on a timescale $\simeq (H/R)/|\zeta-\zeta_L|$ times the
mass transfer timescale ($H$ is the photospheric pressure scale
height; typically $H/R \simeq 10^{-4} - 10^{-2}$)
to very high values. The unstable phase is usually too short to be 
observable. Depending on the system parameters the binary would either
merge, or reappear in a stable semi--detached configuration once the
mass ratio is sufficiently reduced.

\section{The effect of highly relativistic jets}

If the primary black hole is continuously ejecting a highly relativistic jet
powered by the hole's rotational energy, then the energy loss rate in
the jet implies a correponding loss rate $\Gamma \dmej$ of gravitating
mass from the black hole. The quantity $\Gamma$ specifies
the jet energy loss in units of the rest mass $\dmej$ carried in the jet,
i.e.\ can be significantly larger than the Lorentz factor seen in the
bulk motion of the jet.  
As the jet ejection is a consequence of mass accretion, the rest--mass
ejected in the jet must be roughly equal to the rest--mass transferred
from the donor. In addition the jets will presumably carry off the 
specific orbital angular momentum $j_1 = M_2J/M_1M$ of the black hole 
from the binary orbit.
The jet ejection process therefore constitutes a `consequential
angular momentum loss' or CAML process (see e.g.\ King \& Kolb 1995), with
$\dot M_1 = \Gamma \dot M_2 - \dot M_2 = (\Gamma-1)\dot
M_2$ and $\dot J_{\rm jet} = (M_2/M_1) (J/M) \Gamma \dot M_2$. 

The Roche--lobe index for this jet--induced CAML is
\be
    \zeta_L(\Gamma) \simeq \zeta_{L0} - \Gamma \frac{4M_2}{3M},
\label{zetajet}
\ee
assuming $M_2 \la M_1$. If $q=M_2/M_1$ is large the expression becomes
slightly more complicated as ${\rm d}\ln f_2/{\rm d}\ln q \simeq
-1/3(1+q)$ is no longer a good approximation. 
Further details are given in King \& Kolb 1998. 

Thus in the presence of jets with large $\Gamma$ the stability
denominator in (\ref{mdot}) becomes large. This has two
consequences:  
(a)~The mass transfer rate is reduced compared to the case with
conservative mass transfer. 
(b)~Mass transfer is stable even if the mass ratio is large. 
Physically, the jet ejection is equivalent to a massive stellar
wind from the primary which widens the orbit.

Figure~1 shows the quantitative effect of (a) and (b). In the 
left panel we plot the stability term $\zeta-\zeta_L(\Gamma)$, in
the right panel the factor $(\zeta -
\zeta_L(\Gamma)/(\zeta-\zeta_{L0})$ by which the mass transfer rate is
reduced, as a function of mass ratio, for various values of $\Gamma$.
For the example shown we chose $\zeta=0$ and used the full expression for
$\zeta_L(\Gamma)$. The reduction of $(-\dot M_2)$ is largest close to
where conservative mass transfer would be unstable.

\begin{figure}
\vspace*{-0.7cm}
\centerline{\plottwo{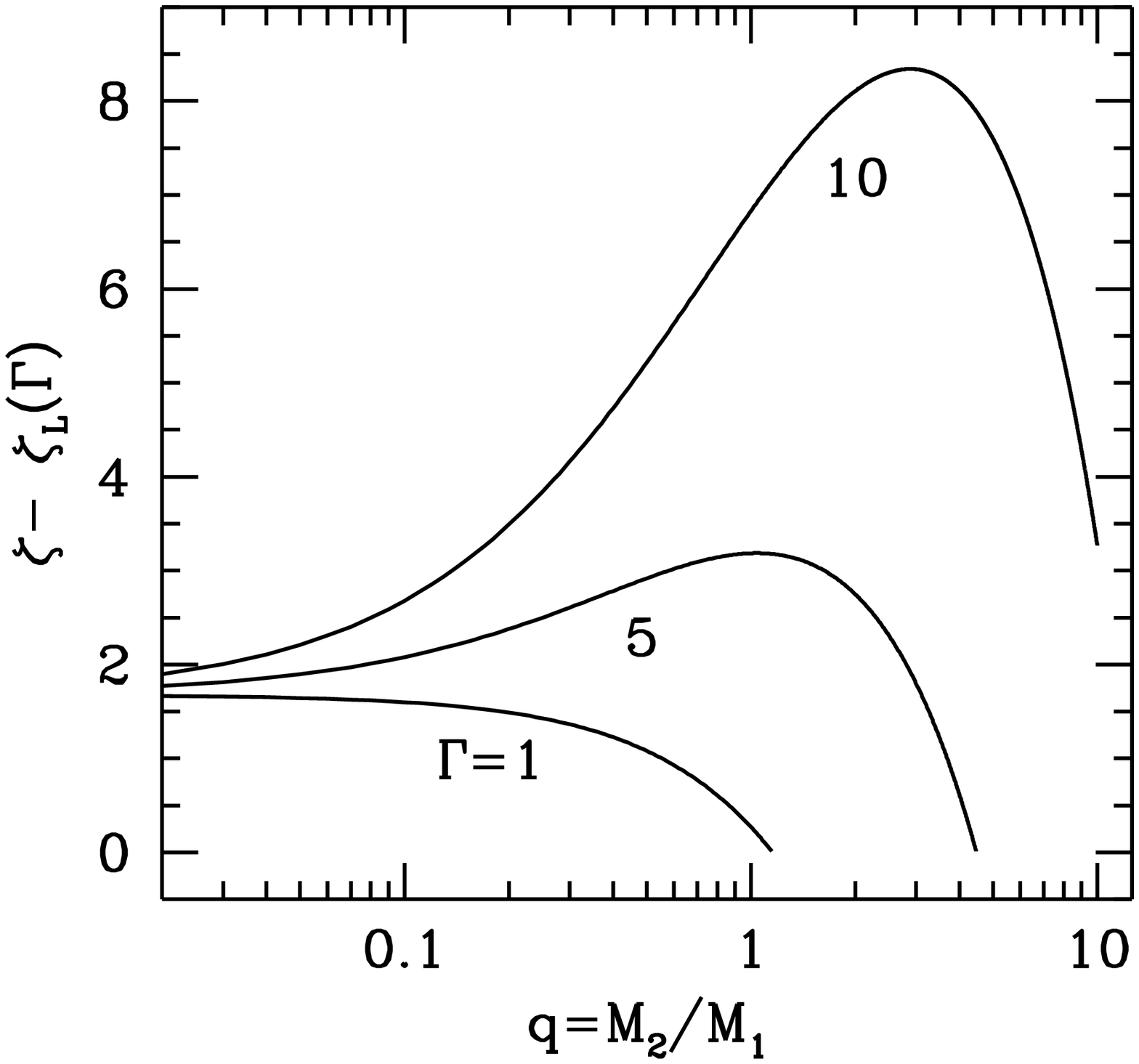}{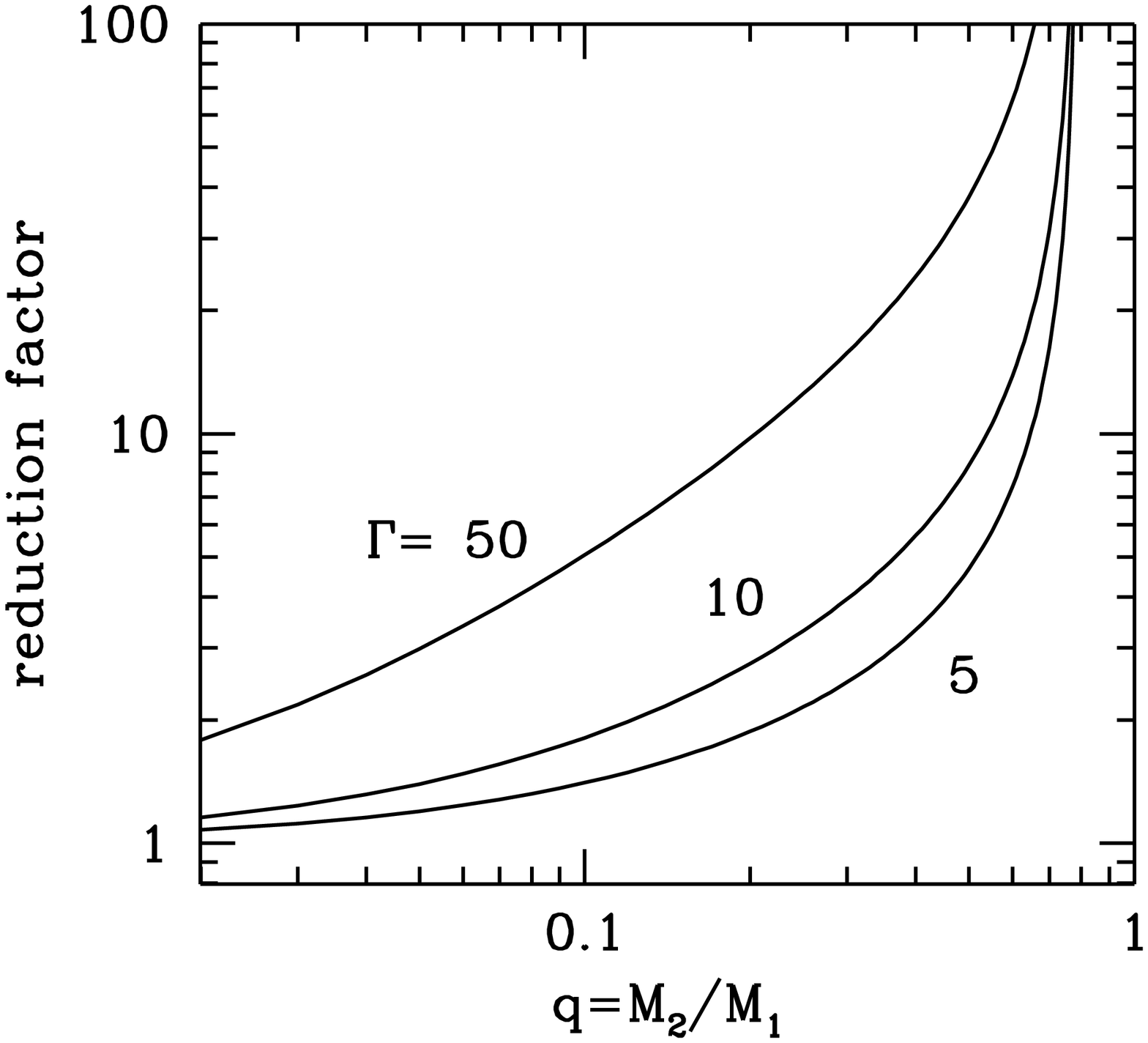}}
\vspace*{-1.0cm}
\caption{Stability term (left) and mass transfer rate reduction factor 
(right) as a function of mass ratio $q=M_2/M_1$ for jets with different
specific energy $\Gamma$.}  
\end{figure}

\section{Discussion}

Jet--induced CAML operates if the jets are powered by the black
hole's rotational energy. The effects on the binary evolution 
are significant only for very energetic jets, but do not depend on 
the detailed mechanism of jet formation and propagation.  

The duration $\Delta t$ of the phase with jet--reduced mass
transfer rate turns out to be largely independent of the jet energy
$\Gamma$:  
The extractable gravitating mass $\Delta M = g(\epsilon) M_1$
from a Kerr black hole with gravitating mass $M_1$
depends on the efficiency $\epsilon$ of the extraction process. In the
case of maximal efficiency the mass can reach the
irreducible value $M_1/\sqrt{2}$, i.e.\ $g(1)=0.29$. For
astrophysically realistic processes (e.g.\ Blandford \& Znajek 1977)
$\epsilon \la 0.5$, in which case $g(\epsilon) \la 0.1$ (King \& Kolb
1998).  Such a limit on $\Delta M_1$ implies a corresponding limit
$\Delta M_{\rm tr} = \Delta M_1/\Gamma$ for the transferred rest--mass
during the jet phase. Hence the duration of the jet phase, $\Delta t
\simeq \Delta M_{\rm tr}/(-\dot M_2)$, is essentially independent of
$\Gamma$, because the mass transfer rate also varies as $1/\Gamma$, see
(\ref{mdot}) and (\ref{zetajet}) for large $\Gamma$. 
We find typically $\Delta t \simeq 0.1 t_{\rm ev}$, where $t_{\rm ev}$ is the
timescale associated with the systemic driving of mass transfer.

Jet--induced CAML would be most pronounced in black
hole systems with massive donor stars. The nuclear expansion of these
is rapid ($K$ in (\ref{mdot}) large). Without jet action such systems
might simply be unobservable, for a number of reasons: \\
1)~The transfer phase is short--lived.\\
2)~The transfer rate is super--Eddington. For an $8 \msun$ black hole 
this is expected if the donor is a main--sequence star with mass 
$\ga 6 \msun$ (case A mass transfer), or if the donor had a mass $\ga
3\msun$ at the end of core hydrogen burning and is now expanding to
the giant branch (case B mass transfer). Mass lost from the binary  
might shroud the system and degrade the X--rays usually expected from
such binaries.\\
3)~The transfer rate is highly super--Eddington. Presumably a common envelope 
forms and the binary merges.

Jet--induced CAML could operate in GRO~J1655-40 and reduce the
transfer rate sufficiently to allow disc instabilities to occur.
The jet would have to be very energetic ($\Gamma
\simeq 50$) to decrease $(-\dot M_2)$ by more than the necessary factor of
10 for the claimed mass ratio $\simeq 0.3$ (Orosz \& Bailyn 1997).   
GRS 1915+105 represents another prime candidate. There are 
indications that the companion star is massive (Mirabel et
al.\ 1997), the jets are very energetic (Mirabel et al.\ 1998), and
the black hole might be close to maximum spin (Zhang et al.\
1997). It is not clear, however, if the system is semi--detached.
SS433 could be affected by jet--induced CAML as well. The nature of
the compact star is still unclear (e.g.\ Zwitter \& Calvani 1989;
D'Odorico et al.\ 1991), but a black hole cannot be ruled out. The
claimed mass ratio is of order 3 or larger, so that the donor is a
massive star if the accretor is indeed a black hole. Again, the donor
might not fill its Roche lobe (Brinkmann et al.\ 1989), and the jets
seem to be less energetic than in the superluminal sources.

Systems that are unstable against conservative mass transfer but
stabilized by jet--induced CAML will still encounter 
the instability at the end of the jet phase. 
The mass loss rate from the hole is much larger than the mass
loss rate from the donor, so that the mass ratio $M_2/M_1$ increases
during the jet phase.

A further consequence of jet--induced CAML is that a large amount of
energy is deposited into the interstellar medium. This can be as much
as $10^{54}$~ergs if $10\%$ of the gravitating mass of
a $10\msun$ Kerr black hole has been extracted at the end of the jet
phase. GRO~J1655--40 would deposit this energy over a time of $\simeq
10^6$~yr. Zhang et al.\ 1997 find that the black hole spin in a number
of black hole X--ray binaries is consistent with zero. It would be
interesting to search for signs of past large energy deposition in the
surroundings of these systems, as they might have been spun down by
jets.  The fact that so far no such effects are observed seems to
indicate that the jet phase terminates much earlier, i.e.\ the black
hole spin probably does not change significantly during the jet
phase. 
   
\smallskip

\noindent {\em Acknowledgements} I thank Andrew King for discussions and
for a careful reading of the manuscript.

\end{document}